\documentclass[aps,prl,superscriptaddress,notitlepage,twocolumn]{revtex4-1}

\usepackage[utf8x]{inputenc}
\usepackage{color}
\usepackage{bbm} 
\usepackage{amsfonts,amsmath,amssymb,stmaryrd}
\usepackage{graphicx}
\usepackage{subfigure}  
\usepackage{hyperref}
\usepackage{epsfig}
\usepackage{mathrsfs}
\usepackage{verbatim}
\usepackage{ulem}

\usepackage{txfonts} 

\renewcommand{\l}{\left(}
\renewcommand{\r}{\right)}

\newcommand{\bra}[1]{\langle#1|}
\newcommand{\ket}[1]{|#1\rangle}

\renewcommand{\a}{\hat{a}}

\newcommand{\ad}{\hat{a}^\dagger}

\newcommand{\Tr}{\text{Tr}}

\usepackage{array}

\definecolor{dgreen}{rgb}{0.0, 0.5, 0.0}

\usepackage{cancel,ifthen}
\newcommand{\cmnt}[2][NoInPuT]{\ifthenelse{\equal{#1}{NoInPuT}}{}{{\color{red}\sout{#1}}} {\color{blue} #2}}

\usepackage{bm}	
\renewcommand{\vec}[1]{\bm{#1}}

\def\qc1{q_c^{(1)}}

\def\gib{g_{\rm IB}}

\newcommand{\veck}{{\vec k}}

\newcommand{\vecp}{{\vec{p}}}
\newcommand{\vecP}{{\vec{P}}}
\newcommand{\vecq}{{\vec{q}}}

\newcommand{\vecQ}{{\vec{Q}}}

\begin{document}

	\normalem	
	
	\title{Dynamical variational approach to Bose polarons at finite temperatures}
	
	\author{David Dzsotjan}
	\affiliation{Department of Physics and Research Center OPTIMAS, University of Kaiserslautern, 67663 Kaiserslautern, Germany}
	\affiliation{Wigner Research Center, Konkoly-Thege ut 29-33, 1121 Budapest, Hungary}
	\author{Richard Schmidt}
	\affiliation{Max-Planck-Institute of Quantum Optics, Hans-Kopfermann-Strasse. 1, 85748 Garching, Germany}
	\affiliation{Munich Center for Quantum Science and Technology (MCQST), Schellingstr. 4, 80799 M\"unchen, Germany}
	\author{Michael Fleischhauer}
	\affiliation{Department of Physics and Research Center OPTIMAS, University of Kaiserslautern, 67663 Kaiserslautern, Germany}
	
	\begin{abstract}
	We discuss the interaction of a mobile quantum impurity with a Bose-Einstein condensate of atoms at finite temperature. To describe the resulting Bose polaron formation we extend the dynamical variational approach  of [Phys.~Rev.~Lett.~\textbf{117}, 11302 (2016)] to an initial thermal gas of Bogoliubov phonons. We study the  polaron formation after switching on the interaction, e.g., by  a radio-frequency (RF) pulse from a non-interacting to an interacting state. To treat also the strongly-interacting regime, interaction terms beyond the Fr\"ohlich model are taken into account. We calculate the real-time impurity Green's function and discuss its temperature dependence. Furthermore we determine the RF absorption spectrum and find  good agreement with recent experimental observations. We predict temperature-induced shifts and a substantial broadening of spectral lines. The analysis of the real-time Green's function reveals a crossover to a linear temperature dependence of the thermal decay rate of Bose polarons as unitary interactions are approached. 
		\end{abstract}

	\date{\today}

	\maketitle
	
\paragraph*{\textbf{Introduction.--}}
The interaction of a mobile impurity with a surrounding quantum bath is one of the paradigmatic models of many-body physics. The polaron introduced by Landau, Pekar and Fr\"ohlich \cite{Pekar,Landau} to describe the motion of 
an electron in a lattice of ions, is formed by the dressing with lattice phonons and is a prime example of quasi-particle formation in condensed matter. More recently neutral atoms immersed in quantum degenerate gases of bosonic or fermionic atoms have attracted much attention since they are experimentally accessible
platforms allowing to study polaron physics with high precision and in novel regimes. Employing Feshbach resonances \cite{Chin2010} it is possible  to tune the impurity-bath interaction from weak to strong coupling
and Rydberg states can be used to study impurities with
non-local interactions \cite{Rydbergpolaron2,SchmidtDem2016,Sous2019}.

\begin{figure}[b!]
		\begin{center}
			\includegraphics[width=0.45\textwidth]{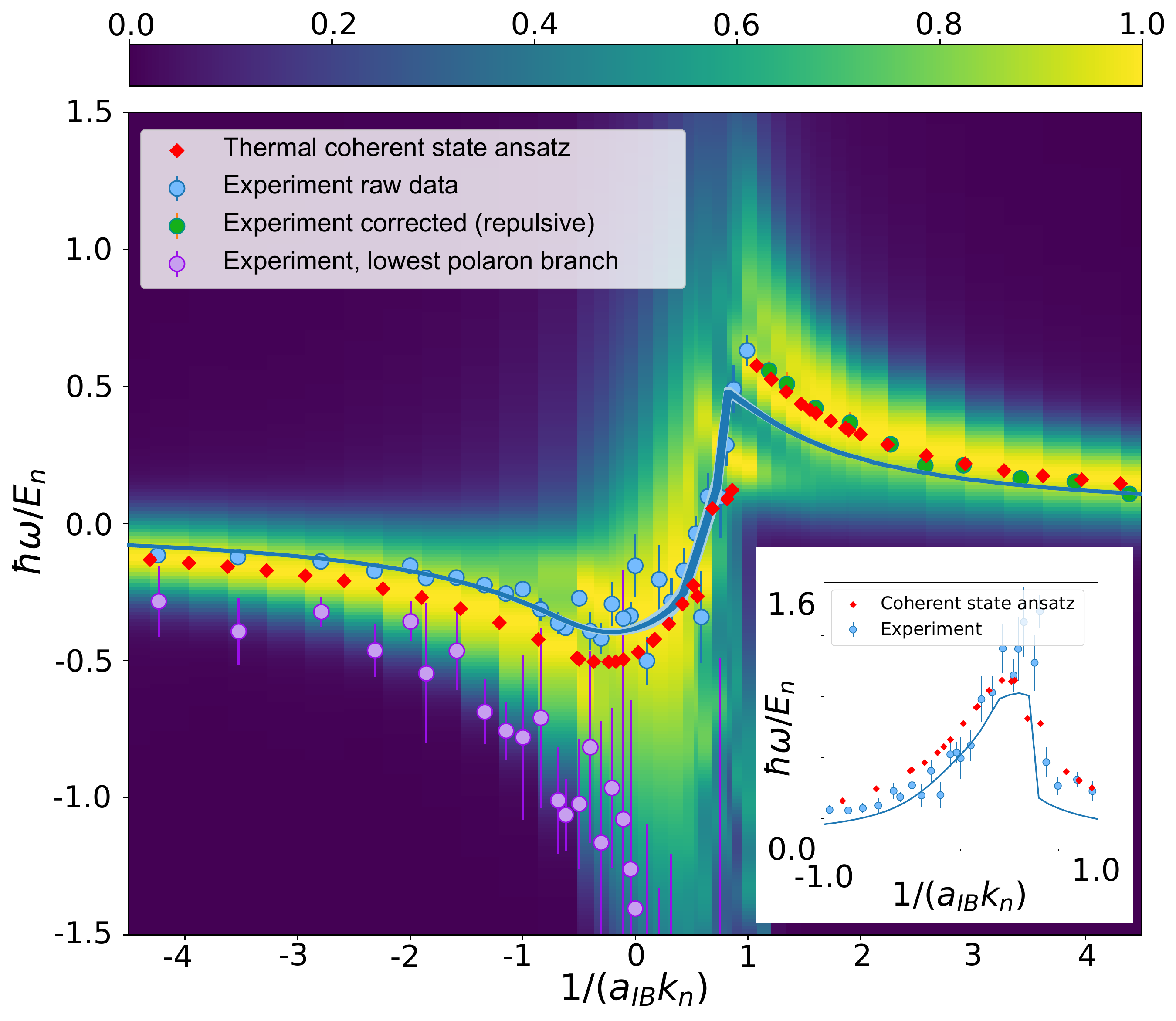}
		\end{center}
		\caption{(Color online) 
		 Density averaged polaron absorption spectra $\overline{A}(\omega)$ from a non-interacting state of the impurity into an interacting   
		 state
		as function of dimensionless impurity-boson scattering length $k_n a_\textrm{IB}$. 
		 $k_n=(6\pi^2 n)^{1/3}$ with $n$ being the trapped-averaged density of the Bose gas (see \cite{jws16}), and $E_n=k_n^2/(2 m_\textrm{red})$.
		 Red dots show the  mean peak values and HWHM width (inset)  of the absorption spectrum predicted by the thermal coherent state variational ansatz (color code) at $T=160\text{nK}$ compared with experimental results (blue and green symbols) from \cite{jws16} and \cite{ardila2018}. Full lines show the $T=0$ theoretical predictions in \cite{jws16}. }
		\label{fig:spectrum}
	\end{figure}

The problem of a Fermi-polaron, i.e. an impurity interacting with a degenerate Fermi gas has been studied in 
a number of experiments in recent years \cite{swa09,Zhang2012,kohstall_metastability_2012,koschorreck_attractive_2012,Scazza2016,cetina_decoherence_2015,cetina_2016,parish2016}. This and related theoretical work \cite{Chevy2006,Lobo2006,Comescot2007,Pilati2008,Punk2009,prok2008,Prokofiev2008b,Mora2009,Kroiss2014,Kroiss2015,goulko2016,schmidt_excitation_2011,
Cui2010,massignan_repulsive_2011,Massignan2014,Schmidt2012b,Ngampruetikorn2012,goold_orthogonality_2011,knap2012,FermiPolaronRevRich,Mistakidis2019} have led to a rather good understanding 
of this problem. In contrast, the description of impurities in a Bose-Einstein condensate (BEC), leading to the so-called Bose polaron is more involved \cite{RathSchmidt2013}. The challenge for theory  is here directly related to the relatively large compressibility of the system, which allows for a much larger number of excitations that can be 
generated by the impurity. Also the experimental observation presented a major challenge due to three-body losses, and has only recently been achieved in experiments at JILA \cite{hvk16}, Aarhus \cite{jws16}, and MIT 
\cite{Yan2019}. Tuning through a Feshbach resonance  all regimes from weak to strong coupling were studied. While being in good general agreement with theoretical predictions, the Aarhus data showed deviations
for strong repulsive interactions, see Fig.~\ref{fig:spectrum}, which were attributed to a nonzero temperature.  Following up on that, a recent extended $T$-matrix analysis predicted rather dramatic temperature effects \cite{Levinsen_FinTPol_2018}, most notably the appearance of new temperature-induced quasi-particle peaks \cite{Guenther2018}.  
These recent developments highlight the need of new theoretical approaches that take into account temperature as well as the creation of a large number of excitations in polaron formation. Indeed, while there exists by now a broad set of theoretical techniques to study Bose polarons at $T=0$,  extensions to $T > 0$ are not straight forward, so that, apart from first diagrammatic and functional determinant approaches \cite{Levinsen_FinTPol_2018,Guenther2018,SchmidtDem2016}, theoretical progress remained highly limited so far.

We here discuss the Bose polaron at non-zero temperature extending the dynamic variational approach of \cite{Shchadilova2016} to an initial $T
>  0$ state of Bogoliubov phonons. The method is a non-equilibrium one and thus gives direct access to the dynamics of polaron formation after an excitation from a non-interacting state by an RF pulse, as it is observed in experiments. One of the key results is depicted in Fig.~\ref{fig:spectrum}, where we show the calculated, trap averaged absorption spectrum $\overline{A}(\omega)$ (color coding) for the Aarhus experiment. For $1/k_na_\textrm{IB} >0$ a notable difference between the peak positions of $T=0$ calculations (full line) from the experimental values (green circles) 
was observed. In contrast, the results from our approach (red points) show good agreement. On the attractive side only uncorrected experimental data for the peak positions is available (blue circles, \cite{jws16}). While this data exhibits a small deviation from our values [we define the mean peak response as $\bar \omega =\int d\omega \omega \overline{A}(\omega)$], the onset of the polaron branch (purple circles, \cite{ardila2018}) matches well the onset of the theoretical absorption spectra. Moreover, the calculated width of the absorption peaks (inset) agrees  well with the experiment. Finally while our approach predicts temperature-induced shifts, thermal quasi-particle broadening, and a temperature-dependent quasi-particle weight, in contrast to \cite{Guenther2018} we do not find evidence for a significant transfer of spectral weight to new quasi-particle peaks.

\paragraph*{\textbf{Model.--} }

We here consider the interaction of a single impurity of mass $M$ with a homogeneous Bose gas in $d$ dimensions in a box of size $L^d$ with periodic boundary conditions. Position and momentum operators of the impurity are  $\hat{\vec{r}}$ and  $\hat {\vec{p}}$.  We treat the BEC of condensate density $n_0$ in Bogoliubov approximation, i.e. in terms of non-interacting plane-wave excitations (phonons) of momentum $\vec{k}$, described by annihilation and creation operators  $\hat a_{\vec{k}}^{(\dagger)}$. The condensate is characterized by the healing length $\xi=1/\sqrt{2 g_\textrm{BB} n_0 m}$, where $m$ is the mass of the BEC atoms and $g_\text{BB}=2\pi a_\textrm{BB} /m$ describes their mutual interaction with $a_\textrm{BB}$ the s-wave scattering length ($\hbar = 1$). The Bogoliubov dispersion relation reads $\omega_\veck= c k\sqrt{1+k^2\xi^2/2}$ with $c= \sqrt{g_\textrm{BB} n_0/m}$ denoting the speed of sound and $k=|\veck|$. In order to formally decouple the impurity we
transform to  a co-moving frame  \cite{Lee1953} using
$\hat U= \exp\bigl\{ - i \hat{\vec{r}} \cdot\hat\vecP_\textrm{ph}\bigr\} $ where $\hat\vecP_\textrm{ph}=
\sum_\veck \vec{k} \hat a_{\vec{k}}^\dagger \hat a_{\veck}$ is the total phonon momentum. Using $
\hat U^\dagger \, \hat\vecp\, \hat U = \hat \vecp - \sum_\veck \veck\, \hat a_\veck^\dagger \hat a_\veck$ and $\hat U^\dagger \hat a_{\vec{k}} \hat U
= \hat a_{\vec{k}}\, e^{-i \vec{k}\cdot\hat {\vec{r}}}$, the Hamiltonian reads \cite{RathSchmidt2013}
\begin{eqnarray}\label{eq:HLLP}
		&& H^\textrm{LLP}(\hat \vecp) =  \frac{1}{2 M}\Bigl(\hat \vecp-\sum_\veck \veck  \ad_{\vec{k}}\a_{\vec{k}}\Bigr)^2 + \sum_\veck \omega_\veck \ad_{\vec{k}} \a_{\vec{k}} + \gib n_0 \nonumber\\
		&&\quad + \frac{g_\textrm{IB}}{L^{d/2}}\sum_\veck  n_0^{1/2} W_\veck \l \a_{\vec{k}} + \ad_{-\vec{k}}  \r \nonumber \\
		&&	\quad +\frac{\gib}{2 L^d}  \sum_{\veck,\veck^\prime} \Bigl[ V^+_{\veck,\veck^\prime} 
		\ad_{\vec{k}} \a_{\vec{k}'} + \frac{1}{2} V^-_{\veck,\veck^\prime} \l \ad_{\vec{k}} \ad_{-\vec{k}'} + \a_{-\vec{k}} \a_{\vec{k}'} \r \Bigr],
\end{eqnarray}
where $V^\pm_{\veck,\veck^\prime} = \l W_\veck W_{\veck'} \pm W_\veck^{-1} W_{\veck'}^{-1} \r$.
$\hat\vecp$ now represents the total momentum of the system which is a constant of motion \cite{Girardeau1961}. In the polaron frame the phonon dynamics attains a nonlinear term $\sim \hat\vecP_\textrm{ph}^2$ which describes impurity-mediated phonon-phonon interactions that vanish in the limit $M\to\infty$. The impurity-BEC interaction strength $g_\textrm{IB}=2\pi a_\textrm{IB}/m_\textrm{red}$ is expressed in terms of the $s$-wave scattering length $a_\textrm{IB}$ and  reduced mass $m_\textrm{red}= m M/(m+M)$, and  $W_\veck =\left[{k^2 \xi^2 }/{ (2 + k^2 \xi^2)}\right]^{1/4}$. 
Crucially note that, due to thermal depletion, the condensate fraction $n_0=n_0(T)$ is temperature dependent. For weak Bose-Bose interactions we have $n_0(T) / n=1 - (T/T_c)^{3/2}$ at fixed total particle density $n$.

Polaron  properties are encoded in the impurity Green's function $ S(t) = \Tr \Bigl\{e^{i H_0 t} e^{- i H t} \rho\Bigr\}$, where the density matrix $\rho$ determines the initial state of the system, and $H_0$ and $H$ are the Hamiltonian in absence and presence of the impurity bath interaction. $S(t)$, also called `dynamical overlap', describes the dephasing dynamics of the system following a sudden quench of $g_\textrm{IB}$ at time $t=0$.  It can be measured using Ramsey spectroscopy as previously demonstrated in fermionic environments \cite{cetina_2016,FermiPolaronRevRich,knap2012,cetina_decoherence_2015}. Fourier transformation of $S(t)$ in turn yields the (injection) absorption spectrum  $A(\omega) = 2 \textrm{Re}\int_0^\infty\!\! d\tau\, e^{i\omega\tau}\, S(\tau)$ in linear response, when the impurity is driven from a non-interacting state to a state with finite $g_\textrm{IB}$ \cite{cetina_2016,FermiPolaronRevRich,schmidt_excitation_2011,massignan_repulsive_2011}.

 In contrast to previous studies of this problem \cite{RathSchmidt2013,Shchadilova2016,mistakidis2018,Camacho2018,Nielsen2019,ardila2018,jws16,Levinsen2015,Sun2017,Ardila2015,will2019,Grusdt2017,Grusdt2018,Ashida2018,lemeshko2016,Midya2016,Enss2019,Volosniev2015}, we here consider finite temperature. This is accounted for by an initial density matrix $ \rho= \rho_\textrm{T}^\textrm{ph} \otimes \ket{\vecp}\bra{\vecp}$ where the phonon bath is in thermal equilibrium, 
$\rho_\textrm{T}^\textrm{ph}= e^{-\beta \sum_\veck \omega_\veck \ad_\veck \a_\veck}/Z$ ($Z=\Tr[e^{-\beta H_0}]$), for an impurity initially in an momentum eigenstate $\ket{\vecp}$. 
In order to unambiguously identify the role of bath temperature and to allow direct comparison with previous studies \cite{Guenther2018} we focus in the following on $\vecp=0$.
While for $T=0$ the initial state is invariant under the
Lee-Low-Pines transformation, this is no longer the case at finite $T$. Introducing a projector $\hat \Pi_\vecQ$ on eigenstates of
$\hat\vecP_\textrm{ph}$ with eigenvalue $\vecQ$ one finds $U^\dagger \rho U
=  \int\! d^3Q \, \hat \Pi_\vecQ\, \rho_\textrm{T}^\textrm{ph} \otimes\ket{\vecp+\vecQ}\bra{\vecp+\vecQ} $.
We represent the thermal state of phonons as a Gaussian average over coherent states $\vert\xi_\veck\rangle $ \cite{ScullyBook}:
$\rho_\textrm{T}^\textrm{ph} =\prod_{\veck} \int d^2\xi_{\vec{k}} \, \frac{ e^{-\vert \xi_{\vec{k}} \vert^2/\bar{n}_{\vec{k}} } } {\pi \bar{n}_{\vec{k}} }\, \, \bigl\vert\xi_{\vec{k}}\bigr\rangle\bigl\langle \xi_{\vec{k}}\big|.$
Here $\bar{n}_{\vec{k}}= 1/(e^{\beta \omega_\veck}-1)$ is the average phonon number in mode $\vec{k}$.
Thus we find 
\begin{eqnarray}\label{StEvolution}
S(t) &=&  \int \!\! d^3  Q \, \overline{\phantom{\Bigl(}\bigl \langle \psi_0(t)\bigr\vert \, \hat \Pi_{\vecQ} \, \bigl\vert \psi(t) \bigr \rangle}.\label{eq:S-var}
\end{eqnarray}
Here $\ket{\psi(t)}= e^{-iH^\textrm{LLP}(\vecQ)t} \vert \xi_\veck\rangle$ and $\ket{\psi_0(t)}=e^{-iH^\textrm{LLP}_0(\vecQ)t} \vert \xi_\veck\rangle$
describe the time-evolution of the initial states $\vert\xi_\veck\rangle$ under $H^\textrm{LLP}(\vecQ)$ 
and $H_0^\textrm{LLP}(\vecQ)$ (Eq.~\eqref{eq:HLLP} for $g_{\text{IB}}=0$), respectively, where the impurity-momentum operator $\hat\vecp$ is replaced by the c-number $\vecQ$.
The overbar denotes the average over the $\xi_\veck$'s
and  we have used that  $H_0^\textrm{LLP}$ commutes with the projector $\hat \Pi_\vecQ$.  
%

\paragraph*{\textbf{Dynamical variational ansatz.--}}
%

%
We calculate $S(t)$ using wave functions $\ket{\psi(t)}$ and $\ket{\psi_0(t)}$ in a variational submanifold of Hilbert space constructed by time-dependent multi-mode coherent states
$\vert \beta(t)\rangle  =\prod_\veck e^{\beta_\veck \hat a^\dagger_\veck-h.c.}\ket{0}=\prod_\veck \vert \beta_\veck(t)\rangle$ including a time-dependent phase $\vert \psi(t) \rangle = e^{-i\phi(t)} \vert \beta(t)\rangle$. With $\vert \psi_0(t) \rangle = e^{-i\phi_0(t)} \vert \beta^0(t)\rangle$, and defining $\Delta\phi\equiv\phi-\phi_0$ one finds
$S(t)  = \int \!\! d^3  Q \, \overline{\phantom{\bigl(} e^{-i\Delta \phi(t)}  \langle \beta^0(t) \vert \, \hat \Pi_\vecQ \, \vert \beta(t)\rangle}.$
The minimization of the Lagrangian ${\cal L}=\bigl\langle\psi(t)\bigr\vert i\partial_t -H^\textrm{LLP}\bigl\vert\psi(t)\bigr\rangle$, and similarly ${\cal L}_0$,  gives the Euler-Lagrange equations $\frac{d}{d t} \left({\partial {\cal L}}/{\partial\dot\beta_\veck}\right)-{\partial {\cal L}}/{\partial\beta_\veck}=0$, 
\begin{eqnarray}
&& i\frac{d}{dt}\beta_\veck(t) = \Bigl(\omega_\veck +\frac{\veck^2}{2M} -\frac{\vec k}{M}\cdot\bigl(\vecQ - \vecP_\textrm{ph}\bigr)\Bigr)\, \beta_\veck(t) 
  + \frac{\gib\sqrt{n_0}}{L^{d/2}} W_\veck\nonumber\\
 &&\qquad+ \frac{\gib}{L^d} \sum_\vecq\Biggl(W_\veck W_\vecq \, \textrm{Re}[\beta_\vecq(t)]+i \, W_\veck^{-1}W_\vecq^{-1}\, \textrm{Im}[\beta_\vecq(t)]\Biggr).\label{eq:Euler-Lagrange-2}
\end{eqnarray}
and similarly for $\beta_\veck^0(t)$ (where $\gib=0$), with (random) initial values $\beta_\veck(0) = \beta_\veck^0(0)=\xi_\veck$. The total phonon momentum in state $\vert\psi(t)\rangle$ is given by
$\vecP_\textrm{ph}=\sum_\veck \veck \vert\beta_\veck(t)\vert^2$. The time-dependent Schr\"odinger equation implies ${\cal L}={\cal L}_0=\text{const}$ and thus
\begin{eqnarray}
i\frac{d \Delta \phi}{dt}  = \gib \biggl(n_0 +\sqrt{\frac{n_0}{L^d}} \sum_\veck W_\veck \beta_\veck^\prime(t) \biggr)-\frac{\bigl(\vecP_\textrm{ph} -\vecP^0_\textrm{ph}\bigr)^2}{2M},\label{eq:Delta-phi}
\end{eqnarray}
with $\beta_\veck^\prime = \textrm{Re}[\beta_\veck]$. Note that different from $T=0$, also the phonon momentum $\vecP^0_\textrm{ph}$ without interactions enters.

\paragraph*{\textbf{Infinitely heavy impurity.--}}

\begin{figure}[t]
		\begin{center}
			\includegraphics[width=0.85\columnwidth]{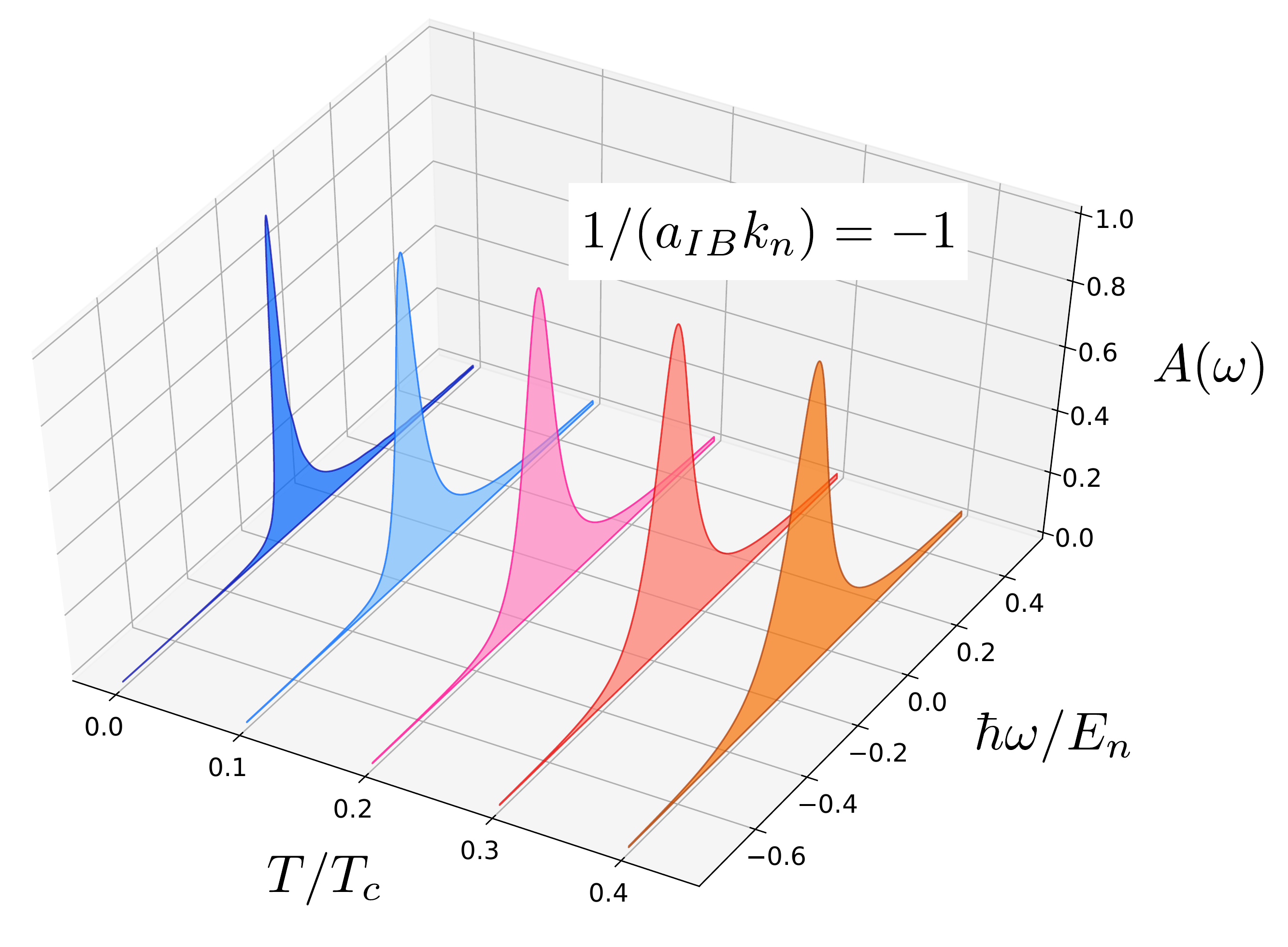}
		\end{center}
		\caption{(Color online)
		Absorption spectrum for attractive polaron for infinitely heavy impurity $M=\infty$ at different
		temperatures, and Bose-Bose interaction strength $ k_na_\textrm{BB} = 0.01$. While increasing $T/T_c$ leads to a shift of the line center and a substantial broadening of the quasiparticle peak, no new quasiparticle peaks appear.		}
		\label{fig:3D-spectraNew}
	\end{figure}

We first discuss the limit of an infinitely heavy impurity, $M\to\infty$. In this case $H_0^\textrm{LLP}(\vecQ)$ and $H^\textrm{LLP}(\vecQ)$ become independent of $\vecQ$ and Eq.~\eqref{eq:S-var} becomes
\begin{equation}
S(t)  = \overline{\phantom{\Bigl(} e^{-i\Delta \phi(t)} \bigl \langle \beta^0(t) \bigr \vert \, \beta(t)\bigr\rangle}.\label{eq:S-inf-M}
\end{equation}
Moreover, the equations of motion (EOM) for $\beta_\veck(t)$, $\beta_\veck^0(t)$, and $\Delta\phi(t)$ become linear. This allows one to express the overlap $\langle \beta^0(t)\vert \, \beta(t)\rangle$ as a matrix-Gaussian function in terms of the random initial variables $\xi_\veck$, and the thermal average can be carried out analytically (see Supplementary Materials). For $g_\textrm{BB}=0$ our coherent state approach becomes exact and we have verified that our results match those from a functional determinant approach \cite{SchmidtDem2016}. Eq.~\eqref{eq:S-inf-M} allows one to determine the temperature dependence of $S(t)$ and from its Fourier transform the absorption spectrum as shown in Fig.~\ref{fig:3D-spectraNew} for a fixed impurity-bath interaction strength. One notices a substantial broadening with increasing temperature accompanied with a small shift of the peak position.

\paragraph*{\textbf{Finite impurity mass.--}}

For a finite impurity mass the variational states are $\vecQ$-dependent and the corresponding integration in Eq.~\eqref{eq:S-var} cannot be carried out upfront. Furthermore the EOM Eqs.~\eqref{eq:Euler-Lagrange-2} become nonlinear  due to the presence of the total phonon momentum $\vecP_\textrm{ph}$.   For $T=0$, ${\vecP}_\textrm{ph}$ is proportional to the conserved polaron momentum and thus vanishes in the case of an impurity initially at rest. This does not hold, however, at finite temperatures where ${\vecP}_\textrm{ph}$ also contains the random initial amplitudes $\xi_\veck$. For this reasons the case of a finite impurity mass is substantially more involved compared to zero temperature even within the coherent-state variational approach and one has to resort to approximations. 

First, the projector on total-momentum eigenstates can be written as $\hat \Pi_Q=\frac{1}{(2\pi)^3}\int \! d^3z\,  e^{i\vec{z}\cdot(\hat \vecP_\textrm{ph}-\vecQ)}$ where the action of the operator $e^{i\vec{z}\cdot\hat \vecP_\textrm{ph}}=\prod_\veck e^{i\vec{z}\cdot\veck \, \hat a_\veck^\dagger \hat a_\veck}$ can be absorbed in a phase shift of the coherent amplitudes $\beta_\veck^0(t)\to \beta_\veck^0(t) e^{-i\vec{z}\cdot\veck}$. Thus evaluating $S(t)$ for a finite impurity mass is formally analogous to the infinite-mass case, however, demanding two additional integrations $\int\! d^3z$ and $\int\! d^3Q$, which presents a numerical challenge. To address this issue, we here replace the phonon-momentum operator in $\hat \Pi_\vecQ$  by its expectation value with respect to the variational wavefunction, followed by an average over the random thermal amplitudes, i.e. $e^{i\vec{z}\cdot \hat \vecP_\textrm{ph}} \to e^{i\vec{z}\cdot\overline{\vecP_\textrm{ph}}}$.
For an impurity initially at rest one has $\overline{\vecP_\textrm{ph}}=0$ and  $\hat \Pi_Q$ becomes the unity operator $\hat \Pi_Q\to \delta^{(3)}(\vecQ-\overline{\vecP_\textrm{ph}})=\delta^{(3)}(\vecQ)$. The $\vecQ$ integration is then trivial and $S(t)$ obeys again Eq.~\eqref{eq:S-inf-M}, where $\beta_\veck(t)$, $\beta^0_\veck(t)$, and $\Delta\phi(t)$ now, however, follow equations~\eqref{eq:Euler-Lagrange-2} and \eqref{eq:Delta-phi} for a finite-mass impurity.

Accordingly, the EOM of  $\beta_\veck(t)$ and $\beta_\veck^0(t)$, which are given by Eq.~\eqref{eq:Euler-Lagrange-2} with $\vecQ=0$, are still nonlinear. As outlined in the Supplementary material, the nonlinear terms in Eqs.~\eqref{eq:Euler-Lagrange-2} and \eqref{eq:Delta-phi} can be approximated by a mean-field ansatz, where the quantities $\vecP_\textrm{ph}\,\beta_\veck(t)$ and $\bigl(\vecP_\textrm{ph}-\vecP^0_\textrm{ph}\bigr)^2$ are effectively replaced by  $ \veck\, \overline{n}_\veck\, \beta_\veck(t)$, and $ 2 \sum_\veck k^2 \, \overline{n}_\veck\bigl(\vert \beta_\veck(t)\vert^2-\vert\xi_\veck\vert^2\bigr)$ respectively.
To this end we note that at $t=0$ the $\beta_\veck(t)$ are  Gaussian random variables given by $\xi_\veck$ and we can assume that they remain Gaussian for all times.
This finally  renders the EOM for  $\beta_\veck(t)$ in a linear form that is amenable to an analytical solution. Note that the equation for the total phase remains nonlinear but can be readily integrated. As a result the dynamical overlap can be expressed as matrix-Gaussian functions in terms of $\xi_\veck$ and  the thermal averaging can be carried out analytically.

\begin{figure}[t]
		\begin{center}
			\includegraphics[width=0.49\textwidth]{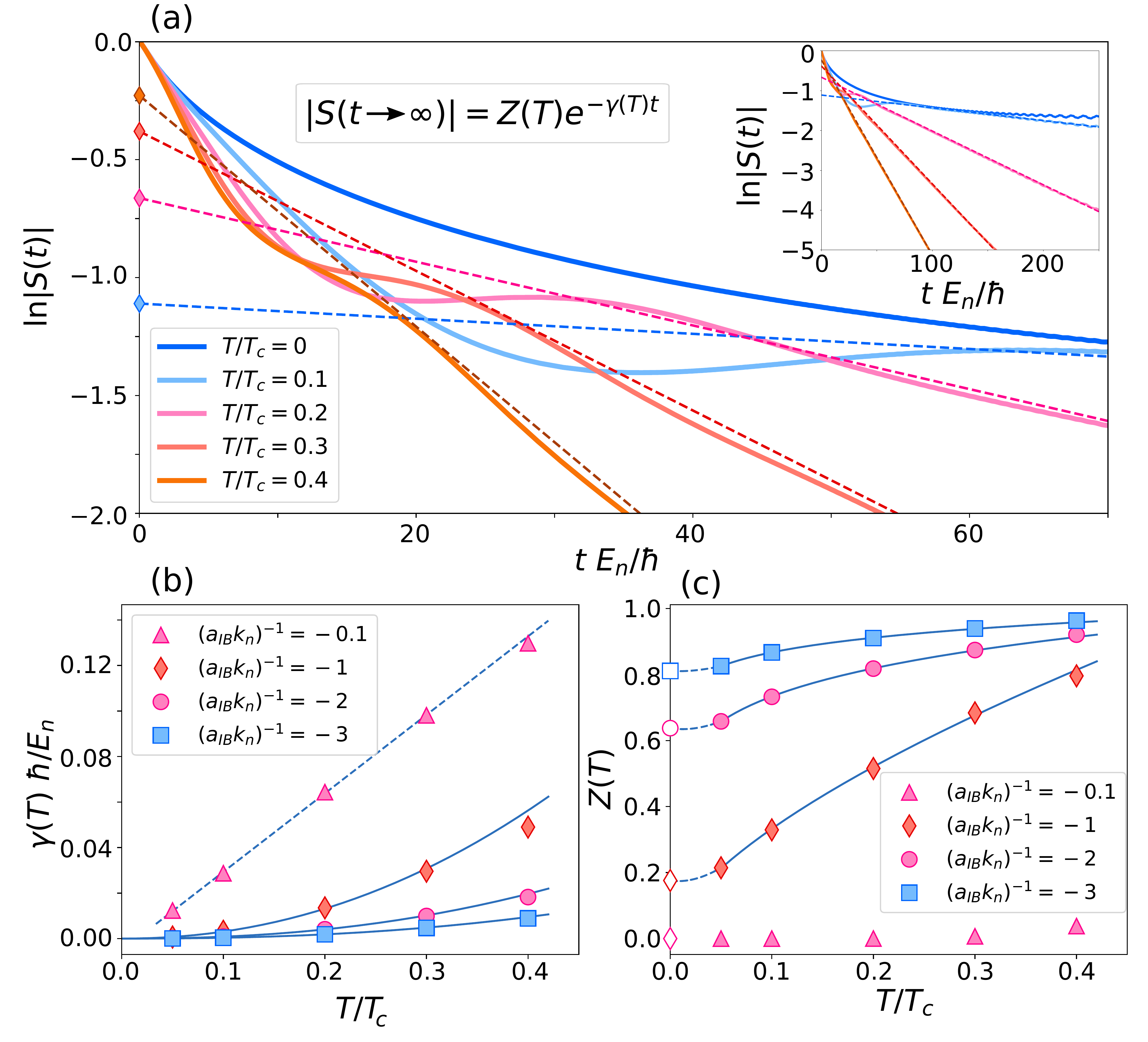}
		\end{center}
		\caption{(Color online) (a)
		Impurity Green's function $S(t)$ of an attractive polaron with $k_n a_\textrm{IB}=-1$ for increasing temperatures for $m=M$ and  $k_n a_\textrm{BB}= 0.01$. While at $T=0$ the overlap $|S(t)|$ approaches a finite value, it turns into an asymptotic exponential decay for $T>0$. Extrapolating the exponential to $t=0$ defines a finite-temperature weight $Z(T)$. (b) Decay rate as function of $T/T_c$ obtained from fits to exponential tails. Close to unitary interactions a linear temperature dependence is found (dashed), linked to quantum critical behaviour in \cite{Yan2019}.  (c) Thermal weights $Z(T)$, where empty symbols show values for $T=0$. As guide to the eye for interaction strength $(k_n a_\textrm{IB})^{-1} = (-1;-2;-3)$
		asymptotic power-law fits for $\gamma(T)\sim (T/T_c)^\nu$ and $Z(T)\sim (T/T_c)^\mu$ are shown with exponents $\nu=(2.10; 2.28; 2.35)$ and $\mu=(0.65; 0.16; 0.07)$.
		}
		\label{fig:dynamical-overlap}
	\end{figure}

The dynamical overlap $S(t)$ calculated within the mean-field approximation is shown in Fig.~\ref{fig:dynamical-overlap}(a) as a function of time. Results are shown for the attractive polaron for increasing temperatures in units of the critical temperature $T_c= (2\pi/m) \bigl(n/\zeta(3/2)\bigr)^{2/3}$ of a non-interacting gas in a box . 
While for zero temperature $S(t)$ approaches a finite value at large times given by the zero-temperature quasiparticle weight, it decays exponentially for $T> 0$ with an asymptotic behavior $\vert S(t)\vert \sim Z(T) \, e^{-\gamma(T)\, t}$.  The decay rates $\gamma(T)$ and thermal weights $Z(T)$ obtained from fits of the asymptotic tails are plotted in Fig.~\ref{fig:dynamical-overlap}(b) and (c) as function of $T/T_c$ for different impurity-Boson interaction strengths (for more details on the analysis of emerging, subleading quasiparticle branches see the Supplementary Materials). 
One recognizes an asymptotic power-law scaling of both quantities as function of $T/T_c$. While this is reminiscent to the fermionic case \cite{SchmidtDem2016}, where the exponents are given by the scattering phase shift at the Fermi momentum, it remains an open question to find analytical expressions for the exponents in the case of Bose polarons where no such a special finite momentum exists and scattering should predominantly take place at small momenta, or a momentum scale $\sim\sqrt{k_BT}$ determined by the thermal de Broglie wave length (at sufficiently large T). Remarkably, we find that close to unitary interactions the broadening of the quasiparticle peak, determined by $\gamma(T)$, shows a crossover to a linear temperature dependence, which may be attributed to quantum critical behavior of impurities  in a Bose gas \cite{Yan2019}.

%

From the Fourier-transform of $S(t)$, such as shown in Fig.~\ref{fig:dynamical-overlap}(a), we have calculated the absorption spectrum for parameters of the experiment of J\o rgensen \textit{et al.} \cite{jws16} and \cite{ardila2018}. The result, shown in Fig.~\ref{fig:spectrum}, is in good agreement with the experiment. The peak positions extracted from the numerical simulations (red points) on the repulsive side coincide with the experimental values (green points) 
determined by Gaussian fits. In order to take into account the inhomogeneous density distribution in the experiment and the finite resolution of the spectrometer we have made a trap average and included a Gaussian broadening using the experimental parameter.  On the attractive side the experimental values for the lowest polaron branch, extracted from the onset of the measured absorption spectrum (purple points), agree with the onset of the absorption spectrum obtained from our numerical simulations.

\paragraph*{\textbf{Summary.--}}

We discussed the physics of a single, mobile quantum impurity interacting with a BEC of atoms at finite temperature. Extending the dynamical variational approach 
of \cite{Shchadilova2016} to the case of an initial thermal state of Bogoliubov phonons, we showed how thermal effects  enter the Hamiltonian in the Lee-Low-Pines frame that is used to decouple the impurity from the phonon dynamics. To describe  polaron formation we calculated the real-time polaron Green's function $S(t)$ and from it the absorption spectrum $A(\omega)$ for a transition of the impurity from a state  non-interacting to a state interacting with the environment.  Strong  impurity-BEC interactions are accounted for by the inclusion of two-phonon terms in the 
 Hamiltonian \cite{RathSchmidt2013}. Within the proposed variational approach one restricts the dynamics to a submanifold of coherent-state wavefunctions that are thermally averaged with Gaussian, random initial amplitudes, with weights determined by the temperature. In this approach observables, such as the the Ramsey signal given by the polaron Green's function are calculated from the solution of nonlinear EOMs for the coherent amplitudes that are subsequently averaged over the thermal distribution. While in the limit of an infinitely heavy impurity the EOMs become linear and the thermal average can be performed analytically, for a finite impurity mass  a mean-field
approximation is required to allow for an analytical thermal average. We calculated the temperature dependence  of  $S(t)$ for different interaction strengths,  and found an asymptotic exponential decay $S(t)\sim Z(T) \, e^{-\gamma(T)\, t}$. The  extracted decay rates $\gamma(T)$ and thermal weights $Z(T)$  show a  power-law dependence on $T/T_c$. Close to unitarity, $1/k_na=0$, the inverse polaron quasiparticle lifetime shows a linear dependence on temperature that is indicative of non-Fermi liquid behavior  in vicinity of the underlying quantum critical point \cite{Yan2019}. The comparison of the theoretical absorption spectra with a recent experiment \cite{jws16,ardila2018} shows that the inclusion of finite temperature corrections leads to excellent agreement between theory and experiment  on the repulsive side that was lacking in previous comparisons with  $T=0$ calculations. Our results are also in excellent agreement with the measured lowest polaron branch as well as the widths of the absorption spectra. In contrast to recent $T$-matrix calculations \cite{Guenther2018} we do not find a splitting or separate temperature-induced
quasi-particle peaks of substantial spectral weight.
 

\subsection*{Acknowledgments}
The authors like to thank N. J{\o}rgensen, J. Arlt for providing the experimental data from \cite{jws16}. We also
thank G. Bruun, E. Demler, Y. Shchadilova, and M. Zwierlein for useful comments and discussions. The work of D.~D. and M.~F. has been supported by the Deutsche Forschungsgemeinschaft
(DFG, German Research Foundation) within the SFB-TR 185 -- 277625399.  R.~S. is supported by the Deutsche Forschungsgemeinschaft (DFG, German Research Foundation) under Germany's Excellence Strategy -- EXC-2111 -- 390814868.

\bibliography{BosePolaronLibrary}

\onecolumngrid
\newpage

\appendix

\begin{center}
\Large{Supplementary Material}
\end{center}


In the following we provide  details about the calculations presented in the main text.

\section{I. Infinite impurity mass}

\subsubsection{A. Dynamical overlap and solution of equations of motion}

In the limit of $M\to\infty$ the multi-mode coherent state dynamical variational ansatz yields the dynamical overlap  given by
Eq.~\eqref{eq:S-inf-M} of the main text. Introducing real and imaginary parts according to  $\beta_\veck = \beta_\veck^\prime + \mathrm{i}\beta_\veck^{\prime\prime}$ 
and $\xi_\veck = \xi_\veck^\prime + \mathrm{i}\xi_\veck^{\prime\prime}$ we find
\begin{eqnarray}
S(t)&=& \overline{\phantom{\Bigl(} e^{-i\Delta \phi(t)} \bigl \langle \beta^0(t) \bigr \vert \, \beta(t)\bigr\rangle}\nonumber\\
 & =& \prod_\mathbf{k}\int\!\!\mathrm{d} \xi^\prime_\veck \int\!\!\mathrm{d} \xi^{\prime\prime}_\veck \frac{1}{\pi \overline{n}_\veck} e^{-\frac{{\xi_\veck^\prime}^2 
 + {\xi_\veck^{\prime\prime}}^2}{\overline{n}_\veck}}e^{-\mathrm{i}\Delta\phi(t)} e^{-\frac{1}{2}\sum_\mathbf{q}|\beta_\vecq(t) - \beta^0_\vecq(t)|^2} 
 e^{\mathrm{i} \sum_\mathbf{q} \textrm{Im}\{\beta_\vecq(t)\beta^0_\vecq(t)^*\}}.\label{eq:S(t)_expanded}
\end{eqnarray}
Here $\beta_\veck(t)$, $\beta^0_\veck(t)$ and $\Delta\phi(t)$ depend on $\xi_\veck^\prime$ and $\xi_\veck^{\prime\prime}$ through the initial conditions of the equations of motion. For convenience, we introduce the variables $\tilde{\beta}_\veck = \sqrt{L^d} \beta_\veck$ and $\tilde{\xi}_\veck = \sqrt{L^d} \xi_\veck$, and omit the tilde from here on. Using the rescaled variables, the EOMs read

\begin{eqnarray}
 \begin{aligned}
  \frac{d}{dt}{\beta}_\veck^\prime & = \omega_\veck \beta_\veck^{\prime\prime} + \frac{g_\textrm{IB}}{L^d}W_\veck^{-1} \sum_\mathbf{q} W_\vecq^{-1} \beta_\vecq^{\prime\prime}, &\quad\quad \beta_\veck^\prime(0) = \xi_\veck^\prime, \\
  \frac{d}{dt}{\beta}_\veck^{\prime\prime} & = -\omega_\veck \beta_\veck^\prime - \frac{g_\textrm{IB}}{L^d}W_\veck \sum_\mathbf{q} W_\vecq \beta_\vecq^\prime - g_\textrm{IB}\sqrt{n_0} W_\veck, &\quad\quad \beta_\veck^{\prime\prime}(0) = \xi_\veck^{\prime\prime},
 \end{aligned}
\end{eqnarray}\label{eq:EOM_beta-infM}
and
\begin{eqnarray}
 \begin{aligned}
  \frac{d}{dt}\Delta{\phi} 
  =  \frac{g_\textrm{IB}\sqrt{n_0}}{L^d}\sum_\mathbf{k} W_\veck \beta_\veck^\prime + g_\textrm{IB}n_0, \qquad\qquad\qquad \quad\Delta\phi(0) = 0.\label{eq:EOM_phi-infM}
 \end{aligned}
\end{eqnarray}
These equations can be expressed in matrix notation,
\begin{equation}\label{eq:mx_eq}
 \dot{\vec{\beta}}(t)=K\, \vec{\beta}(t) + \vec{f},
\end{equation}
where $K$ is a coefficient matrix, $\vec{f}$ is the vector containing the inhomogeneous terms, and the vector $\vec{\beta}$ contains the dynamical variables as
\begin{equation}
 \vec{\beta}(t) = \left(\begin{array}{c}
            \vec{\beta}_\veck^\prime(t) \\
            \vec{\beta}_\veck^{\prime\prime}(t)
           \end{array}\right).\nonumber
\end{equation}
Being an ordinary differential equation, the solution is easily found to be:
\begin{equation}\label{eq:solution1}
 \vec{\beta}(t) = e^{Kt}\left(\vec{\xi} + K^{-1} \vec{f}\right) - K^{-1} \vec{f},
\end{equation}
where the initial condition $\vec{\beta}(0)=\vec{\xi}$ was used. For the following calculations the coefficient matrix ${{K}}$ has to be diagonalized. Since $K$ is non-hermitian this requires the determination of  the right and left eigenvectors which can be done numerically, to obtain the diagonal matrix $\Lambda = U_L K U_R$, where $U_L$ and $U_R$ contain the left and right eigenvectors of $K$, respectively. It is convenient to reexpress Eq.~\eqref{eq:solution1} in the compact form
\begin{equation}
 \vec{\beta}(t)= A(t)\vec{\xi}+(A(t)-\mathbf{1})\vec{w}, 
\end{equation}
where we introduced $A(t) = e^{K t}$  and $\vec{w} = K^{-1}\vec{f}$. This expression is used in Eq.~\eqref{eq:EOM_phi-infM} which after integration over time yields
\begin{eqnarray}
 \begin{aligned}
  \Delta\phi(t) &= g_\textrm{IB}\sqrt{n_0} \sum_\mathbf{q}  W_\vecq\left[B(t)\vec{\xi}+(B(t)-\mathbf{1}\cdot t)\vec{w}\right]_\vecq + c_0(t),
 \end{aligned}
\end{eqnarray}
where $B(t) = U_R (e^{\Lambda t} - \mathbf{1} )\Lambda^{-1}U_L $ and $c_0(t)=g_\textrm{IB} n_0 t$.

The matrices and vectors can be written in separate blocks,%
\begin{eqnarray*}
 \begin{aligned}
  A(t)&=\left(\begin{array}{cc}
                C(t) & D(t) \\
                E(t) & F(t)
              \end{array}\right), \quad \quad \quad B(t)=\left(\begin{array}{cc}
                                                                            G(t) & H(t) \\
                                                                            I(t) & J(t)
                                                                          \end{array}\right), \quad\quad \vec{w}=\left(\begin{array}{c}
                                                                                                            \vec{x}_\veck \\
                                                                                                            \vec{y}_\veck
                                                                                                          \end{array}\right),
 \end{aligned}
\end{eqnarray*}
which allows one to derive the formal solution for the real and imaginary parts of $\beta_\veck$:
\begin{eqnarray}\label{eq:sliced_dynamics}
 \begin{aligned}
  \vec{\beta}^\prime(t)&=C(t){\vec{\xi}^\prime}+D(t)\vec{\xi}^{\prime\prime}+(C(t)-\mathbf{1})\vec{x}+D(t)\vec{y}, \\
  \vec{\beta}^{\prime\prime}(t)&=E(t){\vec{\xi}^\prime}+F(t)\vec{\xi}^{\prime\prime}+E(t)\vec{x}+(F(t)-\mathbf{1})\vec{y}, \\
  \Delta\phi(t)&=g_\textrm{IB}\sqrt{n_0}\sum_\mathbf{q} W_\vecq\left[G(t){\vec{\xi}^\prime}+H(t)\vec{\xi}^{\prime\prime}+(G(t)-\mathbf{1}\cdot t)\vec{x}+H(t)\vec{y}\right]_\vecq + c_0(t).
 \end{aligned}
\end{eqnarray}
%

\subsubsection{B. Thermal average}

Having derived the explicit solutions of the EOMs, we can now calculate the dynamical overlap at finite temperature. To this end the thermal average, encoded in the Gaussian integrations in Eq.~\eqref{eq:S(t)_expanded}, has to be performed.

Since we assume $\textbf{p}\rightarrow 0$ for the impurity momentum, the problem becomes spherically symmetric. We can thus perform an angular average of $\beta_\veck(t)$ and $\xi_\veck(t)$ and only the dependence on the radial component of $\veck$ remains. Discretizing the radial momentum component of $\veck$ as $k_q$ with $q=1...N_\veck$ one can write
\begin{equation}
 \frac{1}{L^3}\sum_\mathbf{k} \rightarrow \sum_{q=1}^{N_k} \frac{\Delta k}{2\pi^2} k_q^2.
\end{equation}
Here $\Delta k$ denotes the difference between the discretized radial $k$ values. Taking this into account, and substituting the solutions \eqref{eq:sliced_dynamics} for the rescaled $\beta_\veck^\prime$, $\beta_\veck^{\prime\prime}$ and for $\Delta\phi$,  the exponents in Eq.\eqref{eq:S(t)_expanded} can be written in a vectorised form. Grouping them according to terms quadratic, linear and constant in $\mathbf{\xi}$, one  obtains
\begin{eqnarray}
 \begin{aligned}
  -\mathrm{i} \Delta\phi(t) &= -\mathrm{i}\vec{\rho}^I \vec{\xi} + -\mathrm{i} \kappa, \\
  -\frac{1}{2}\frac{1}{L^3}\sum_{\vec{k}} |\beta_\veck(t) - \beta_\veck^0(t)|^2 &= - \frac{1}{2}\vec{\xi}^T M^{II}_1 \vec{\xi} + \vec{\rho}^{II} \vec{\xi} - \frac{1}{2}\vec{w}^T M^{II}_2 \vec{w}, \\
  \mathrm{i}\frac{1}{L^3}\sum_{\vec{k}} \mathrm{Im}\{\beta_\veck(t) \beta_\veck^0(t)\} &= \mathrm{i} \vec{\xi}^T M^{III}_1 \vec{\xi}  + \vec{\rho}^{III} \vec{\xi}.  
 \end{aligned}
\end{eqnarray}
The detailed form of the  matrices $M$ follows directly from the substitution of the solutions \eqref{eq:sliced_dynamics}. Also the Gaussian probability distribution can be written in a vectorised form,
\begin{equation}
 \prod_\mathbf{k}\int\!\!\mathrm{d} \xi^\prime_\veck \int\!\!\mathrm{d} \xi^{\prime\prime}_\veck \frac{1}{\pi \overline{n}_k} e^{-\frac{{\xi_\veck^\prime}^2 
 + {\xi_\veck^{\prime\prime}}^2}{\overline{n}_k}} = \int\mathrm{d}^{2N_k}\xi\,\,\sqrt{\frac{\det{\Gamma}}{(2\pi)^{2N_k}}}\,\, e^{-\frac{1}{2}\vec{\xi}^T \Gamma \vec{\xi}},\nonumber
\end{equation}
where $\Gamma$ is a $2N_k \times 2N_k$ diagonal matrix, containing the average phonon occupation numbers, $\overline{n}_k$ as 
\begin{equation}
 \Gamma = \frac{\Delta k}{\pi^2}\left( \begin{array}{cc}
         \begin{array}{ccc}
         k_1^2/\, \overline{n}_1 & \, & \, \\
         \, & \ddots & \, \\
         \, & \, & k_{N_k}^2/\, \overline{n}_{N_k}
         \end{array} & 0 \\
            0 & \begin{array}{ccc}
                 k_1^2/\, \overline{n}_1 & \, & \, \\
         \, & \ddots & \, \\
         \, & \, & k_{N_k}^2/\, \overline{n}_{N_k}
                \end{array}
        \end{array}\right).\nonumber
\end{equation}
Rewriting Eq.\eqref{eq:S(t)_expanded} in terms of these vectorized expressions, one obtains a simple, $2N_k$ - dimensional Gaussian integral:
\begin{equation}\label{eqFDAExtended}
 S(t) = \int\mathrm{d}^{2N_\veck}\xi\,\,\sqrt{\frac{\det{\Gamma}}{(2\pi)^{2N_\veck}}}\,\, e^{-\frac{1}{2}\vec{\xi}^T (\Gamma + M_T)\vec{\xi}}\,\, e^{ \vec{\rho}\cdot\vec{\xi}}\,\, e^{ -\frac{1}{2}\vec{w}^TM_0\vec{w} }\,\,  e^{-\mathrm{i}\kappa},
\end{equation}
where $M_T  = M^{II}_1 + \mathrm{i} (M^{III} + (M^{III})^\top)$, $M_0 = M^{II}_1$, $\vec{\rho} = \vec{\rho}^{II} + \mathrm{i}(\vec{\rho}^{III} - \vec{\rho}^I)$, and $\kappa$ is a phase term originating from $\Delta\phi(t)$.
The integral can be performed analytically. Importantly, however, the form Eq.~\eqref{eqFDAExtended} allows for an  instructive physical interpretation in terms of the different scattering processes mediated by the impurity that contribute to the many-body dynamics of the system. In fact, the exponent quadratic in $\vec{\xi}$ represents scattering of excitations from the thermal reservoir back into the thermal reservoir. The linear-$\vec{\xi}$ exponent contains the contribution of scattering of excitations from the BEC into the thermal part and vice versa, while the third exponent expresses the scattering processes from the BEC back to the BEC. At zero temperature, only the latter term survives while above the critical condensation temperature $T_c$, only the first term is present. This shows that our model indeed encompasses all possible channels of scattering processes in the system which on its own is an important finding.  In particular, it is the second term that allows the conversion between atoms in the condensate fraction and the thermal contribution in BECs at finite temperatures. In previous approaches using functional determinants to, e.g., describe Rydberg Bose polarons in finite temperature BECs \cite{SchmidtDem2016} this contribution was not accounted for and the present approach shows not only how it can be derived explicitly, but it also provides a form amenable to numerical evaluation.

Performing the Gaussian integrals, we obtain a closed, analytic expression for the dynamical overlap:
\begin{equation}
 S(t) = \frac{e^{ \frac{1}{2}\vec{\rho}^T(\Gamma + M_T)^{-1}\vec{\rho} }}{\sqrt{\det(\mathbf{1} + \Gamma^{-1} M_T)}}\,\,e^{ -\frac{1}{2}\vec{w}^TM_0\vec{w} }\,\,  e^{-\mathrm{i}\kappa}.
\end{equation}
Besides the advantage that this result enables one to forego any numerical sampling and averaging, it also has a very clear structure: the finite-temperature dynamical overlap can be written as the product of a $T=0$ part and a finite-temperature factor. Hence the absorption spectrum, obtained by Fourier transformation becomes a simple convolution integral. As $T \rightarrow 0$, due to Bose-statistics,  the occupation of finite-momentum states goes to zero. As a result, $\Gamma^{-1} \rightarrow 0$, thus the finite-temperature part disappears, leaving 
\begin{equation}
 \lim_{T\rightarrow 0}S(t) = e^{ -\frac{1}{2}\vec{w}^TM_0\vec{w} }\,\,  e^{-\mathrm{i}\kappa},\nonumber
\end{equation}
which means we are indeed left with the impurity-mediated scattering processes from the BEC back into the BEC, and a result that recovers the previous findings at zero temperature \cite{Shchadilova2016}.

\section{II. Finite impurity mass and mean-field approximation}

For a finite impurity mass $M$ and $T>0$, the Lee-Low-Pines transformation acting on the initial density matrix makes the initial state of the impurity effectively dependent on the total phonon momentum (according to the relation $U^\dagger \rho U
=  \int\! d^3Q \, \hat \Pi_\vecQ\, \rho_\textrm{T}^\textrm{ph} \otimes\ket{\vecp+\vecQ}\bra{\vecp+\vecQ} $ stated in the main text), which has nonzero fluctuations in a thermal state. Consequently, the variational wavefunctions must be evaluated for all finite values $\vecQ$  and the dynamical overlap is subsequently obtained as the average over $\vecQ$ (see Eq.~\eqref{eq:S-var} in the main text),
\begin{equation}
S(t)  = \int \!\! d^3  Q \, \overline{\phantom{\Bigl(} e^{-i\Delta \phi(t)} \bigl \langle \beta^0(t) \bigr \vert \, \hat \Pi_\vecQ \, \bigl \vert \beta(t)\bigr\rangle}
= \frac{1}{(2\pi)^3}\int \!\! d^3  Q \int\!\! d^3 z\, \, \overline{\phantom{\Bigl(} e^{-i\Delta \phi(t)} \bigl \langle \beta^0(t) \bigr \vert \, e^{-i(\hat{\vecP}_\textrm{ph}-\vecQ)\cdot\vec{z}}
\, \bigl \vert \beta(t)\bigr\rangle}.
\end{equation}
Making use of $e^{i\hat{\vecP}_\textrm{ph}\cdot\vec{z}}\vert \beta_\veck^0(t)\rangle = \vert e^{i\veck\cdot\vec{z}} \beta_\veck^0(t)\rangle\equiv \vert \tilde\beta^0(z,t)\rangle$ this expression can again be expressed in terms of coherent-state overlaps
\begin{equation}
S(t)  = \frac{1}{(2\pi)^3}\int \!\! d^3  Q \int\!\! d^3 z\, \, e^{i\vecQ\cdot\vec{z}} \,
\overline{\phantom{\Bigl(} e^{-i\Delta \phi(t)} \bigl \langle \tilde\beta^0(z,t) 
\, \bigl \vert \beta(t)\bigr\rangle}.
\end{equation}
%
To deal with the challenge that the double integration in this expression prohibits a direct analytical approach to the problem,  we employ here a mean-field approximation which replaces the phonon momentum operator in the expression $ e^{-i\hat{\vecP}_\textrm{ph}\cdot\vec{z}}$ by its expectation value in the variational wavefunctions, averaged over the random initial values, i.e. $ e^{-i\overline{\vecP_\textrm{ph}}\cdot\vec{z}}$. Since $\overline{\vecP_\textrm{ph}}$
is proportional to the conserved total momentum, for an impurity initially at rest it holds $\overline{\vecP_\textrm{ph}}=0$. With this the $\vecQ$ and $\vec{z}$ integrations can be performed explicitly
and we obtain an expression that is formally identical to the dynamical overlap in the $M\to\infty$ case, i.e.
\begin{equation}
S(t)= \overline{\phantom{\Bigl(} e^{-i\Delta \phi(t)} \bigl \langle \beta^0(t) \bigr \vert \, \beta(t)\bigr\rangle}.\label{eq:S-nl}
\end{equation}
Note, however, that unlike the infinite mass case,  the EOMs determining the time-evolution of the coherent amplitudes and phases are now nonlinear equations
\begin{eqnarray}
 i\frac{d}{dt}\beta_\veck(t) = \Bigl(\omega_\veck +\frac{\veck^2}{2M} +\frac{\vec k}{M}\cdot\vecP_\textrm{ph}\Bigr)\, \beta_\veck(t) 
  + \frac{\gib\sqrt{n_0}}{L^{d/2}} W_\veck
 + \frac{\gib}{L^d} \sum_\vecq\Biggl(W_\veck W_\vecq \, \textrm{Re}[\beta_\vecq(t)]+i \, W_\veck^{-1}W_\vecq^{-1}\, \textrm{Im}[\beta_\vecq(t)]\Biggr),\label{eq:EOM-beta-nl}
\end{eqnarray}
\begin{eqnarray}
i\frac{d}{dt} \Delta \phi(t) = \gib n_0 +\gib\sqrt{\frac{n_0}{L^d}} \sum_\veck W_\veck \textrm{Re}[\beta_\veck(t)] 
-\frac{1}{2 M}\left(\vecP_\textrm{ph} -\vecP^0_\textrm{ph}\right)^2,\label{eq:EOM-phi-nl}
\end{eqnarray}
 owing to the fact that they still contain the total phonon momentum $\vecP_\textrm{ph}=\sum_\vecq\, \vecq\,  \beta_\vecq^*(t)\beta_\vecq(t)$ which is a random variable due to the initial values of $\beta_\veck$ and $\beta_\veck^0$. 

In order to treat these nonlinearities in the EOMs we make further approximations: First one notes that for an impurity initially at rest, the total phonon momentum vanishes when thermally averaged over the stochastic initial values, $\overline{\vecP_\textrm{ph}}=0$. Consequently, when  replacing the phonon momentum by its average over the stochastic initial values, the nonlinear terms in Eqs.~\eqref{eq:EOM-phi-nl} and \eqref{eq:EOM-beta-nl} would disappear. This would however be a too strong approximation and we proceed instead as follows:
At $t=0$ the coherent amplitudes $\beta_\vecq(t=0)=\xi_\vecq$ are indeed Gaussian random variables.  Assuming that they remain approximately Gaussian with respect to the average over the $\xi_\vecq$ variables one can apply a mean-field factorization ($\delta x \equiv x- \overline{x}$)
\begin{eqnarray*}
\delta x\,  \delta y\,  \delta z&=&\overline{\delta x\delta y}\,\delta z+\overline{\delta x\delta z}\, \delta y+\overline{\delta y\delta z}\,\delta x\\
\delta w\, \delta x\,  \delta y\, \delta z &=& \overline{\delta w\delta x}\,  \delta y\, \delta z + \delta w\, \delta x\, \overline{\delta y\delta z} -  \overline{\delta w\delta x} \, \overline{\delta y\delta z} \\
& +&\overline{\delta w\delta y}\,  \delta x\,\delta z + \delta w\,\delta y\, \overline{\delta x\delta z} -  \overline{\delta w\delta y} \, \overline{\delta x\delta z} \\
&+ &\overline{\delta w\delta z} \, \delta x\, \delta y + \delta w\, \delta z\, \overline{\delta x\delta y} -  \overline{\delta w\delta z} \, \overline{\delta x\delta y}.
\end{eqnarray*}
In the present case, for an impurity initially at rest one has $\sum_\vecq \vecq \overline{\beta_\vecq^*\beta_\vecq} =0$. Moreover, there is only a coupling between coherent amplitudes of different modes if the two-phonon terms are significant. Neglecting this cross coupling one finds 
\begin{eqnarray*}
\overline{\beta_\vecq^*(t)\beta_\veck(t)} -\overline{\beta_\vecq^*(t)}\,\overline{\beta_\veck(t)} &=& \overline{\xi_\vecq^*\xi_\veck} -\overline{\xi_\vecq^*}\,\overline{\xi_\veck} =
\delta_{\vecq\veck} \overline{n}_\veck,\\
\overline{\beta_\vecq(t)\beta_\veck(t)} -\overline{\beta_\vecq(t)}\,\overline{\beta_\veck(t)} &=& \overline{\xi_\vecq\xi_\veck} -\overline{\xi_\vecq}\,\overline{\xi_\veck} =0,
\end{eqnarray*}
as well as
\begin{eqnarray*}
\overline{\beta_\veck^0} = 0,\qquad \overline{\beta_\vecq^0(t)\beta_\veck^0(t)}  =0,\qquad \overline{\beta_\vecq^{0*}(t)\beta_\veck^0(t)} = \overline{\xi_\vecq^*\xi_\veck} =
\delta_{\vecq\veck} \overline{n}_\veck.
\end{eqnarray*}
Furthermore, recognizing that 
\begin{eqnarray*}
\overline{\beta_\vecq(t)\beta_\veck^0(t)}  =0,\qquad \overline{\beta_\vecq^{*}(t)\beta_\veck^0(t)} = \overline{\xi_\vecq^*\xi_\veck} =
\delta_{\vecq\veck} \overline{n}_\veck,
\end{eqnarray*}
%
	\begin{figure}[b]
		\begin{center}
			\includegraphics[width=0.4\textwidth]{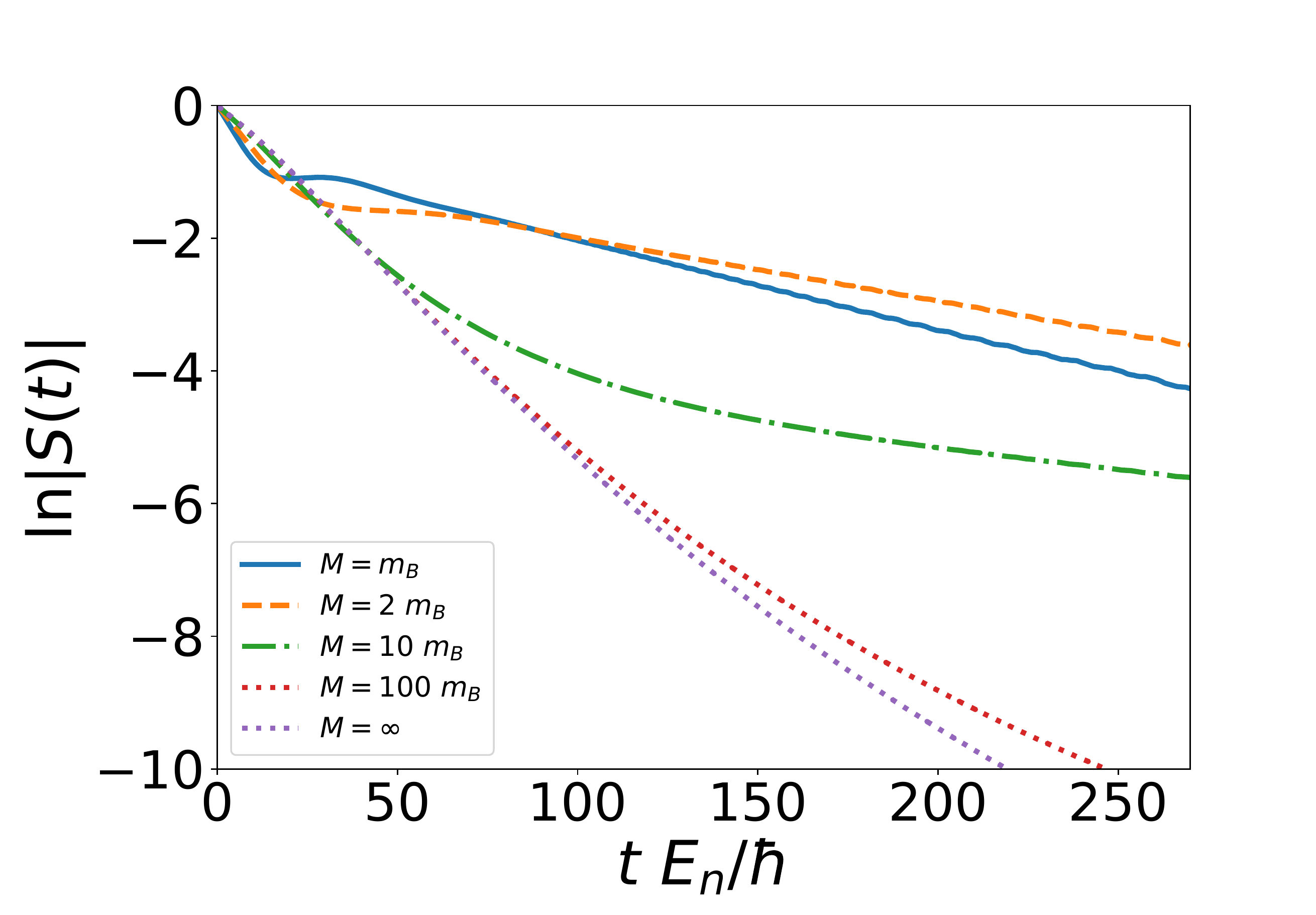}
		\end{center}
		\caption{
		Absolute value of the dynamical overlap for an attractive polaron at $T/T_c = 0.2$ for different impurity masses, a weakly interacting condensate $k_n a_\textrm{BB}= 0.01$, and an impurity-bath interaction strength $1/k_n a_\textrm{IB} = -1$. The initial rapid decrease of $|S(t)|$ turns  into an exponential decay at long times. At the given interaction parameters for an infinitely heavy impurity fitting the asymptotic tails  becomes numerically  challenging since it  requires the calculation of the evolution up to extremely long times at consequently small values of $|S(t)|$.}
		\label{fig:dynamical-overlap-infM}
	\end{figure}
one is finally led to the following mean-field approximation of the nonlinear terms in the EOMs
\begin{eqnarray*}
 \vecP_\textrm{ph}\,\beta_\veck(t)=\sum_\vecq \vecq \beta_\vecq^*(t)\beta_\vecq(t) \beta_\veck(t) \approx \sum_\vecq \vecq \biggl(\overline{\beta_\vecq^*\beta_\vecq}\, \beta_\veck +(\overline{\beta_\vecq^*\beta_\veck}-\overline{\beta_\vecq^*}\,\overline{\beta_\veck}) \, \beta_\vecq +(\overline{\beta_\vecq\beta_\veck}-\overline{\beta_\vecq}\,\overline{\beta_\veck})\, \beta_\vecq^*\biggr)\approx 
 \veck\, \overline{n}_\veck\, \beta_\veck(t),
\end{eqnarray*}
and 
\begin{eqnarray*}
&&\Bigl(\vecP_\textrm{ph}-\vecP_\textrm{ph}^0\Bigr)^2 =\sum_{\veck\vecq} \, \veck\, \vecq \,
\Bigl(\vert\beta_\veck(t)\vert^2 
- \vert\beta_\veck^0(t)\vert^2 \Bigr)\Bigl(\vert\beta_\vecq(t)\vert^2 
- \vert\beta_\vecq^0(t)\vert^2 \Bigr)\\
&&\quad\approx 2 \sum_{\veck\vecq} \, \veck\, \vecq \, \biggl( \Bigl(\, \overline{\beta_\vecq(t)^*\beta_\veck(t)} -\overline{\beta_\vecq(t)^*}\, \overline{\beta_\veck(t)}\,\Bigr)
\, \beta_\vecq(t)\beta_\veck^*(t)
+\overline{\beta_\vecq(t)^{0*}\beta_\veck^0(t)}\,  \beta_\vecq^0(t)\beta_\veck^{0*}(t)
-\overline{\beta_\vecq(t)^{0*}\beta_\veck(t)} \, \beta_\vecq^0(t)\beta_\veck^{*}(t)
-\overline{\beta_\vecq(t)^{*}\beta_\veck^0(t)} \, \beta_\vecq(t)\beta_\veck^{0*}(t)\,\biggr)\\
&&\quad= 2 \sum_\veck \veck^2 \overline{n}_\veck\, \Bigl(\, \vert\beta_\veck(t)\vert^2 - \vert\xi_\veck\vert^2\Bigr).
\end{eqnarray*}
	\begin{figure}[t]
		\begin{center}
			\includegraphics[width=0.65\textwidth]{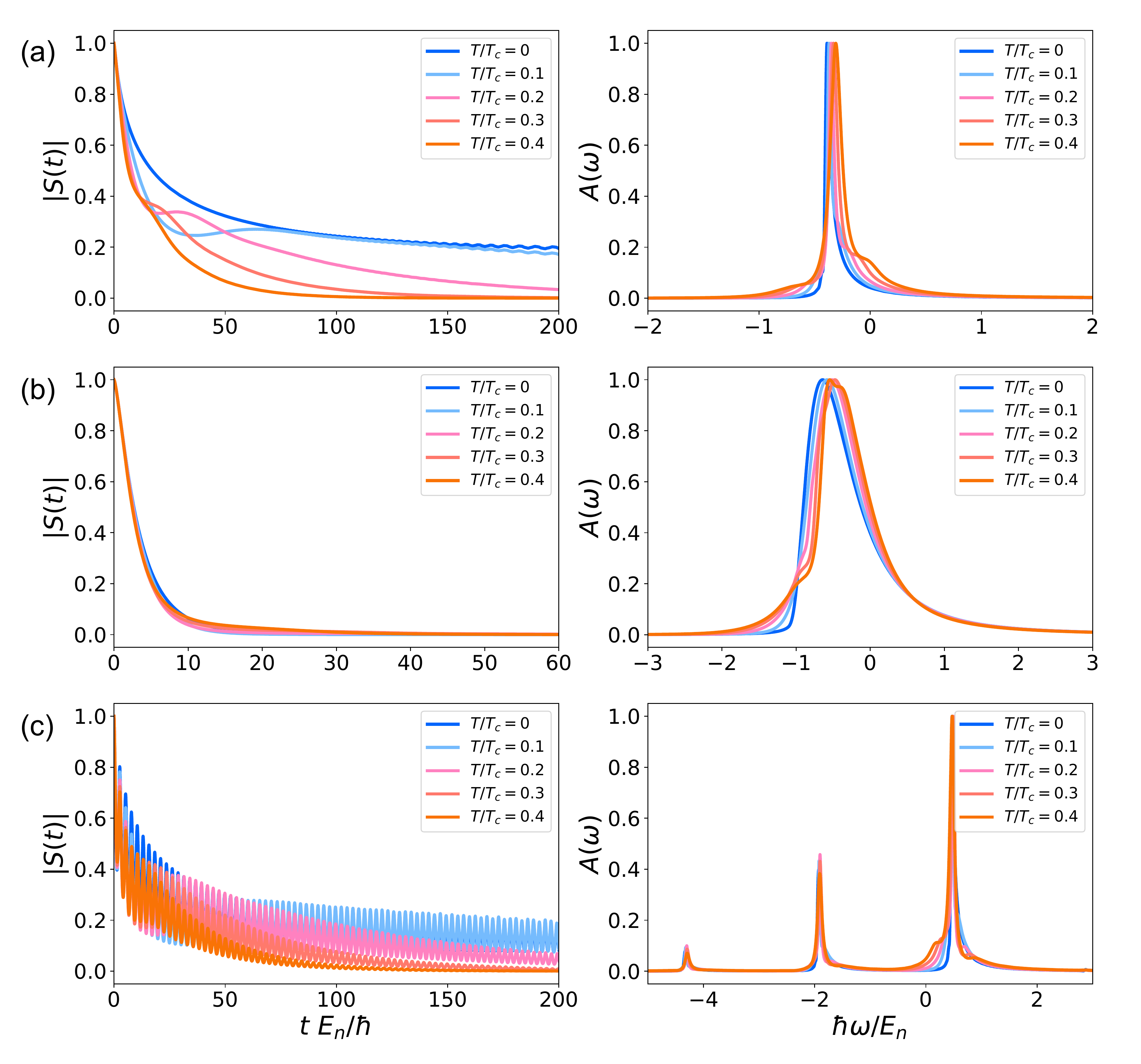}
		\end{center}
		\caption{
		Overlap $S(t)$, experimentally measurable as Ramsey contrast, and reverse absorption spectrum $A(\omega)$ for $m=M$ and $k_n a_\textrm{BB}=0.01$ as function of temperature for attractive [$1/k_na_\textrm{IB}=-1$] (a), strongly attractive [$1/k_na_\textrm{IB}=-0.3$] (b), and repulsive [$1/k_na_\textrm{IB}=1$] (c), interactions. }
		\label{fig:spectrum-cuts}
	\end{figure}
These expressions give, after insertion into the EOM for $\beta_\veck$ the \textit{linear} equations
\begin{eqnarray}
i\frac{d}{dt}\beta_\veck(t) &=& \Bigl(\omega_\veck +\frac{\veck^2}{2M}(1+2 \overline{n}_\veck)\Bigr)\, \beta_\veck(t) 
  + \frac{\gib\sqrt{n_0}}{L^{d/2}} W_\veck
 + \frac{\gib}{L^d} \sum_\vecq\Biggl(W_\veck W_\vecq \, \textrm{Re}[\beta_\vecq(t)]+i \, W_\veck^{-1}W_\vecq^{-1}\, \textrm{Im}[\beta_\vecq(t)]\Biggr),\label{eq:EOM-beta-nl}
 \end{eqnarray}
 %
 %
 \begin{eqnarray}
i\frac{d}{dt} \Delta \phi(t) &=& \gib n_0 +\frac{1}{M}\sum_\veck \veck^2 \overline{n}_\veck\, \Bigl[\vert\beta_\veck(t)\vert^2-\vert\xi_\veck\vert^2\Bigr]
+\gib\sqrt{\frac{n_0}{L^d}} \sum_\veck W_\veck \textrm{Re}[\beta_\veck(t)].
\end{eqnarray}
When two-phonon terms are not important, we can further replace $\vert\beta_\veck(t)\vert^2$ in the second term of the latter equation by the solution of Eqs.\eqref{eq:EOM-beta-nl} without mode-mixing two-phonon terms
\begin{equation}
\frac{d}{dt}\beta_\veck(t) \approx -i \Bigl(\omega_\veck +\frac{\veck^2}{2M}(1+2 \overline{n}_\veck)\Bigr)\, \beta_\veck(t) 
  -i \frac{\gib\sqrt{n_0}}{L^{d/2}} W_\veck
=
 -i\Omega_\veck \beta_\veck(t) -ig_\textrm{IB} \sqrt{\frac{n_0}{L^d}} W_k,\nonumber
\end{equation}
which reads
\begin{eqnarray*}
\beta_\veck(t) = -i \frac{g_\textrm{IB} \sqrt{\frac{n_0}{L^d}} W_k}{\Omega_k}\left(1-e^{-i\Omega_k t}\right) +\xi_\veck e^{-i\Omega_k t}= \overline{\beta}_\veck(t) +\xi_\veck e^{-i\Omega_k t}.
\end{eqnarray*}
With this the solution of the EOMs for $\beta_\veck(t)$ and $\Delta\phi(t)$ have the same form as the corresponding equations in the $M\to\infty$ limit and hence all further calculations can be performed in a similar way as outlined for the case of an  infinitely heavy impurity.  

In order to illustrate the impact of a finite impurity mass and the related  recoil the impurity experiences in collisions with bath excitations in Fig.~\ref{fig:dynamical-overlap-infM} we show the dependence of the impurity Green's functions $S(t)$ on the mass ratio $M/m_B$ at fixed temperature $T/T_c=0.2$ and interaction strength $1/k_na_\textrm{IB}=-1$. In Fig.~\ref{fig:spectrum-cuts} we give examples for the dynamical overlap and the absorption spectra for different temperatures for the cases of attractive, strongly attractive and repulsive interactions. The $T=0$ results agree with those from Ref.\cite{Shchadilova2016}.

\section{III. Spectral analysis of the time-dependent Green's function S(t)}

	\begin{figure}[h!]
		\begin{center}
			\includegraphics[width=0.77\textwidth]{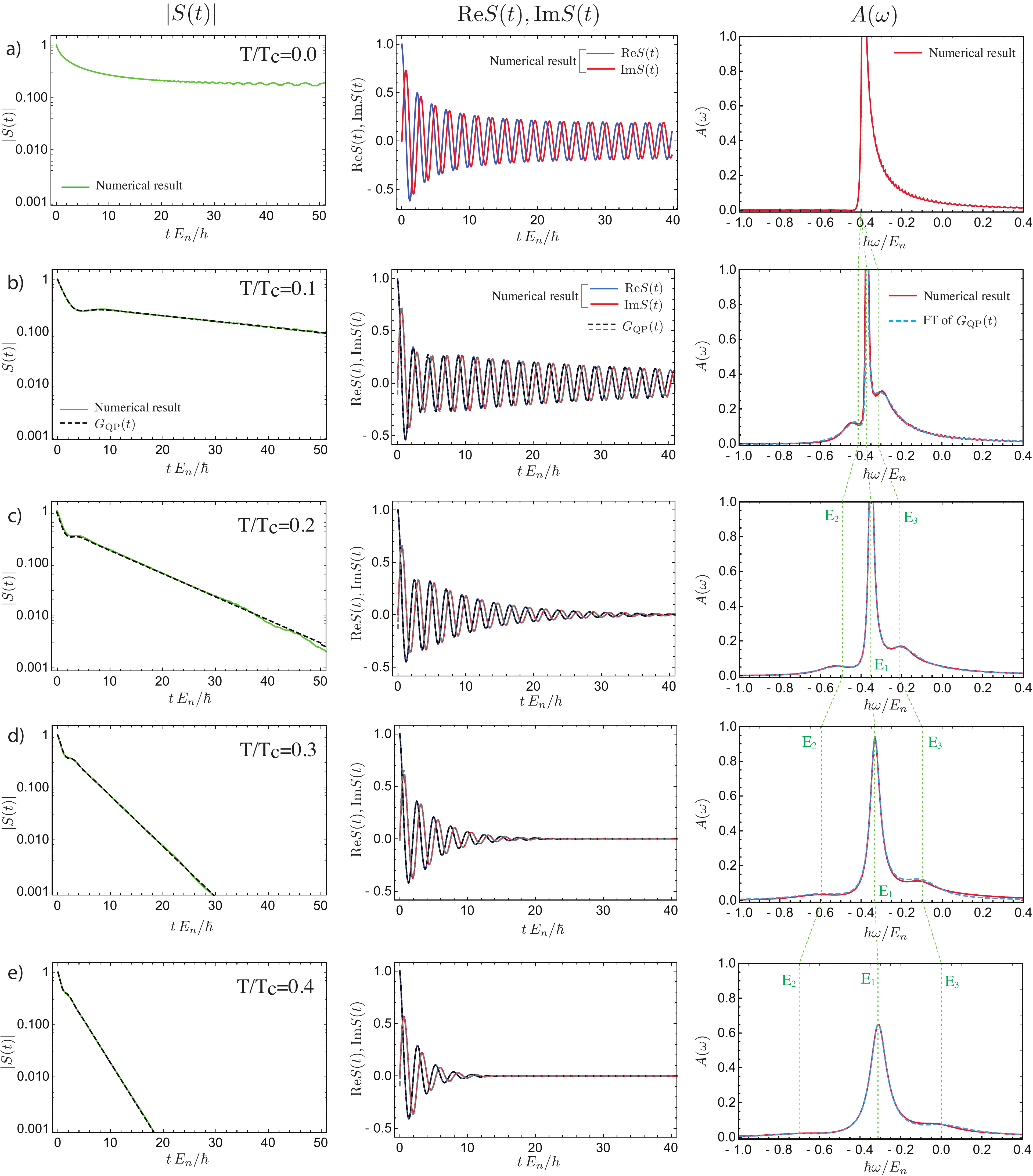}  
		\end{center}
		\caption{Quasiparticle analysis of the impurity Green's function $S(t)$ (solid curves in left and middle column) using fits of $S(t)$ by the quasi-particle approximation $G_\text{QP}(t)$, Eq.~\eqref{QPApprox} (dashed curves), for increasing temperatures (rows (a)-(e)). The right column shows the spectral function resulting from a  Fourier transform of the numerical data of $S(t)$ (red solid), compared to the analytical direct Fourier transformation of Eq.~\eqref{QPApprox}. Indicated as vertical lines are the extracted quasiparticle energies $E_i$. Model parameters are $1/k_na_\textrm{IB} = -1$, $m/M = 1$, and $k_n a_\text{BB}=0.01$. }
		\label{fig_appAna}
	\end{figure}

The real-time Green's function $S(t)$ and the corresponding spectral function $A(\omega)$, such as shown in Fig.~\ref{fig:spectrum-cuts}, contain  information about possible quasi-particle excitations (here at zero momentum), including their energy, lifetime, and quasiparticle weights. In a regime where quasiparticle excitations dominate the spectral response,  the full time-resolved Green's function $S(t)$ may be approximated as a sum over such quasiparticle contributions
\begin{equation}
S(t)\approx G_\text{QP}(t)\equiv \sum_i^{N_\text{QP}} Z_i e^{-i E_i t} e^{-\gamma_i t} e^{-i \phi_i}.\label{QPApprox}
\end{equation}
Here $E_i$ and $1/\gamma_i$ represent  quasiparticle energies and  lifetimes, respectively, and the quasiparticle weights are given by $0\leq Z_i\leq 1$. The $\phi_i$ account for a possible phase arising due to overlapping spectral weights.

In Ref.~\cite{Guenther2018} is was predicted using a diagrammatic approximation that at finite temperature $T$ the attractive polaron peak splits into two quasiparticle excitations which share the weight of the original ($T=0$) polaron peak. Our approach does not predict such a splitting. By fitting the approximation \eqref{QPApprox} to the full numerical data of $S(t)$, we analyse the frequency-resolved properties of the impurity Green's function and we find that, while the polaron peak shifts energetically, it does pertain and even \textit{increase} its quasiparticle weight as temperature increases,  contrary to the prediction in Ref.~\cite{Guenther2018}.

Importantly, the analysis of the numerically obtained spectral function (see Fig.~\ref{fig:spectrum-cuts}) shows that, for mobile impurities, weight is accumulated in emergent, \textit{additional}  quasiparticle-like spectral features that accompany polaron excitations present at $T=0$.  As an example we show in Fig.~\ref{fig_appAna} the results of fitting Eq.~\eqref{QPApprox} with $N_\text{QP}=3$ to the numerical data $S(t)$ in the strongly interacting regime $k_n a_\textrm{IB} = -1$. In each row of Fig.~\ref{fig_appAna} we show from the left to right as solid curves the magnitude of $S(t)$, its real and imaginary part, and the spectral function $A(\omega)$ resulting from Fourier transformation. The columns in turn correspond to increasing temperature from $T/T_c = 0$ to $T/T_c = 0.4$. For finite $T$ the quasiparticle fit function $G_\text{QP}(t)$ is applied to the data and the resulting fits are shown as dashed lines. 

It is evident that not only the absolute value $|S(t)|$ (left column) is extremely well reproduced by the fit but even the full complex function $S(t)$ from short to long times (middle column). Consequently, also the spectra are fit well by the simple approximation \eqref{QPApprox} of the full impurity Green's function. Moreover, we find that the fits using $G_\text{QP}(t)$ yield lifetimes, quasiparticle weights and energies that are fully consistent with the simpler single quasiparticle fit at long times used in the analysis  discussed in the main text (see corresponding Fig.~\ref{fig:dynamical-overlap}), supporting the robustness of our analysis. 

	\begin{figure}[h!]
		\begin{center}
			\includegraphics[width=0.42\textwidth]{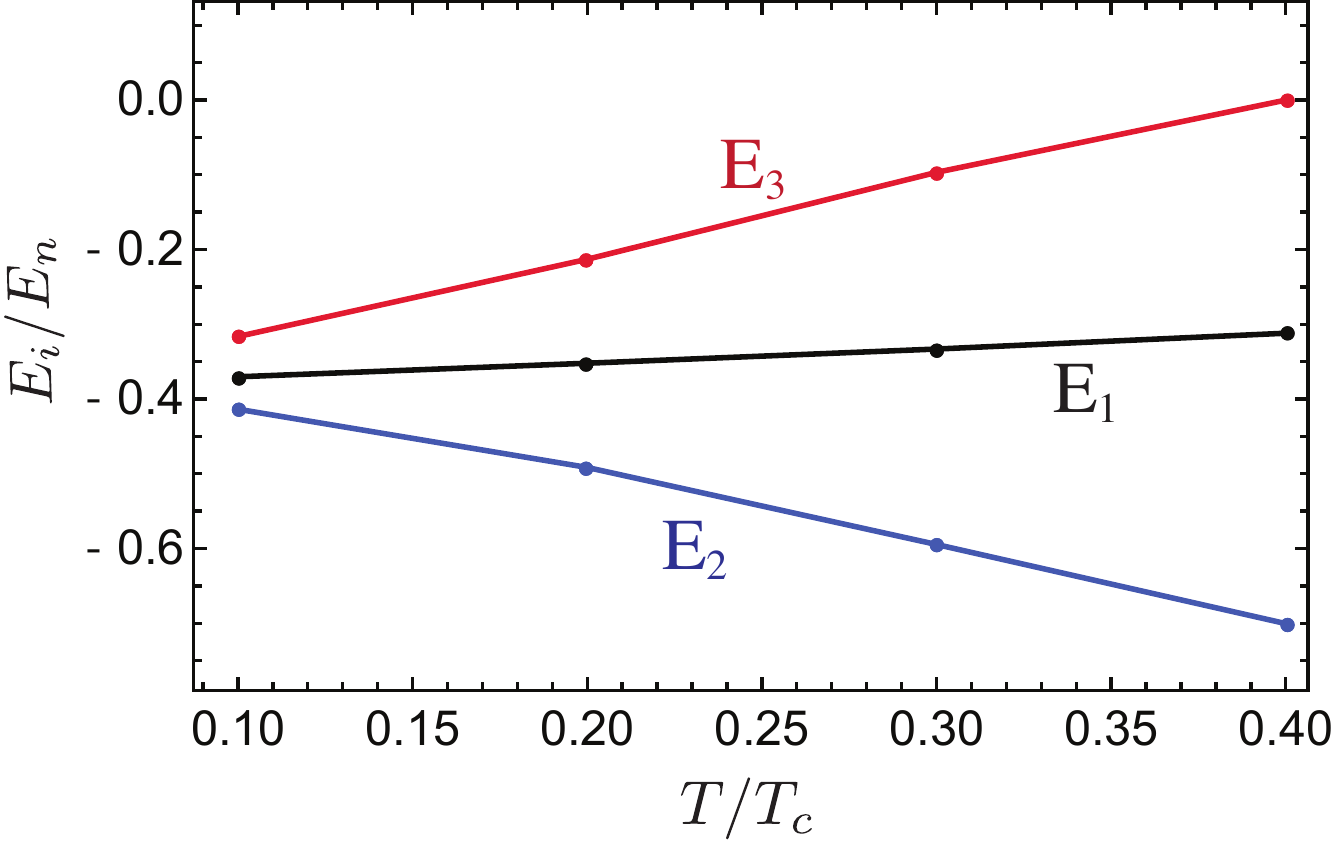}  
		\end{center}
		\caption{Temperature dependence of quasiparticle energies $E_i$ (symbols) for $1/k_na_\textrm{IB} = -1$, $m/M = 1$, and $k_n a_\text{BB}=0.01$. The energies are extracted from a fit of the quasiparticle approximation, Eq.~\eqref{QPApprox} to the data of $S(t)$ numerically obtained using the variational thermal coherent state approach. Solid lines are guide to the eye.}
		\label{fig_appQPEnergy}
	\end{figure}

The analysis in Fig.~\ref{fig_appAna} also not only reveals  how the polaron energy $E_1 = E_\text{pol}$ shifts with $T/T_c$ (Fig.~\ref{fig_appQPEnergy}, c.f. also the vertical, dashed lines in the right column in Fig.~\ref{fig_appAna}), but also helps to identify the temperature-dependence of the two additional spectral features at $E_{2,3}$ accompanying the dominant polaron peak at $E_1$. Indeed we find that while the features at energies $E_{2,3}$ do not absorb weight from the dominant polaron peak, they do become increasingly well defined for increasing temperature and show a clear energetic separation from the main peak at energy $E_1$ (Fig.~\ref{fig_appQPEnergy}). 

We emphasize that the Fourier transform of $S(t)$ yields the frequency resolved (retarded) Green's function $G(\omega)$. Hence, our analysis also gives insight into the approximate pole structure of $G(\omega)$, which can, to a good approximation, be described by a small number of poles in the complex frequency plane with increasingly larger residues as temperature increases.  It remains an interesting and open question whether these features are robust under more advanced variational wavefunction manifolds that, e.g., include squeezing terms beyond the LLP and Bogoliubov unitary transformations included in our approach, and whether the emergent spectral features can be efficiently occupied in an adiabatic preparation in experiments while avoiding three-body losses.

\end{document}